\documentclass[journal]{IEEEtran}
\usepackage{graphicx}
\usepackage[utf8]{inputenc} 
\usepackage[T1]{fontenc} 
\usepackage{amsmath}
\usepackage{amsmath}
\usepackage{caption}
\usepackage{multirow}
\usepackage[autostyle=true]{csquotes}

\begin{document}
	\title{Voltage Quality Time Series Classification using Convolutional Neural Network}
	\author{Sagnik~Basumallik,~\IEEEmembership{Student Member,~IEEE,}}

\maketitle

\begin{abstract}
This paper presents the effectiveness of convolutional neural network (CNN) to classify power quality problems. These problems arise mainly due to increase in use of non-linear loads, operation of devices like adjustable speed drives and power factor correction capacitors, which is a growing concern both for utilities and customers. This work uses the advances in supervised learning to classify different power quality time-series waveforms such as voltage sag, swell, interruption, harmonics, transients and flicker. CNN results in a very high classification accuracy compared to other traditional and machine learning methods in presence of noise. This process can be employed by utilities as well as customers to understand the cause and mitigate voltage quality problems. 
\end{abstract}


\IEEEpeerreviewmaketitle

\section{Introduction}

The major causes of degradation of power quality are due to increase in use of non-linear loads, capacitors and loads switching events, transformer energizing and faults. These events often introduces harmonics and also result in voltage and current sags, swells, interruptions and flickers. Significant economic loss resulting from voltage and current quality degradation have lead to development of standards such as IEEE 1159, which set the guidelines and recommends practices for monitoring power quality problems \cite{intro1}. Timely classification of such events are of utmost importance to understand their impact on costly power system equipments and sensitive loads. Disturbances created by non-linear loads affect distribution system equipments, electric drives, programmable logic controllers, and single-loop controllers \cite{intro2}. A timely identification and cause of such disturbances can help system operators take necessary actions to prevent device malfunctions. \\

A large body of work has been devoted to extract features and classification of such signals. These method includes traditional techniques such as Fourier Transform, Short-Time Fourier transform, Gabor transform, S-transform, Wavelet Transform and Kalman Filter, as well as more recent machine learning methods such as support vector machines and  artificial neural networks \cite{lit1}, \cite{lit2}. In this paper, we extend the work done by authors in \cite{powerQ_CNN} to classify more events such as interruptions, harmonics, transients and flickers using CNN. In order to verify the correctness of our method, signals obtained from RTDS are used (will be) to validate the learning model. 

\section{Power Quality Events}

\begin{figure*}
	\centering
	\includegraphics[scale=0.3]{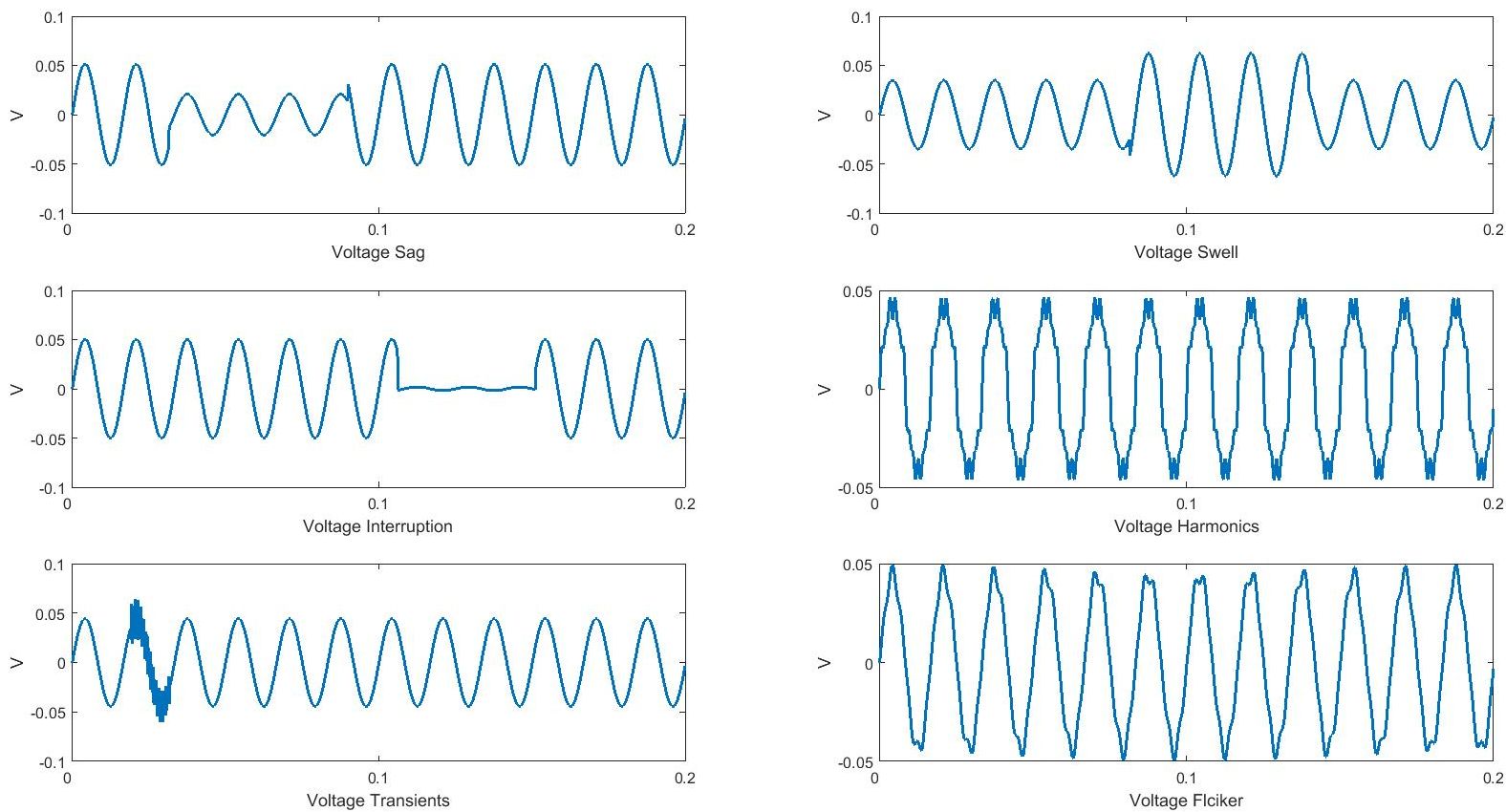}
	\caption{Normalized voltage waveforms for six different power quality events}
	\label{fig:plot_all}
\end{figure*}

Different events in power system have different influences on the quality of power being delivered. Such events can be either caused by faults or switching operations. In case of a line-to-line or line-to-ground fault, voltage interruptions or sags can be observed. Operations like shunt switching, induction motor starting, transformer energizing and non-linear load switching result in voltage notches, swells, and sags, flickers, besides introducing harmonics \cite{classifiy1}. Thus, classifying these events are important to analyze the behavior of power systems under different operating conditions. Six different power quality scenarios are studied in this paper. The following section explain in brief the different events along with their mathematical formulations. Fig. \ref{fig:plot_all} shows the various power quality events under the study. 
\subsection{Voltage Sag}

Voltage sags, which fall under the category of `under-voltage', occurs mainly during heavy load switching or power system faults. IEEE Standard 1159 describes this condition as the decrease of 10\%-90\% of RMS voltage for a time duration between 0.5 cycles to 1 min. Voltage sags can also be caused as a result of inductor motor starting, transformer energizing or increase in source impedance \cite{classify2,classify3}. It is also termed as `votage dip' and can be mathematically described by (\ref{sag}) where $0.1 \leq \alpha \leq 0.9$,\begin{equation}\label{sag}
v(t)=V[1-\alpha(u(t-t_1)-u(t-t_2))]sin(\omega t)
\end{equation}

\subsection{Voltage Swell}

Voltage swell fall under the category of `over-voltage' which occurs when the voltage increase by 10\% to 90\% of its RMS value for 0.5 cycles to 1 min. Such events are caused due to sudden reduction in load, switching a capacitor bank, temporary voltage rise on unfaulted phase during SLG fault \cite{classify2,classify3}. It is also termed as `monemtary overvoltage' and is mathematically given by (\ref{swell}) where $0.1 \leq \alpha \leq 0.9$, \begin{equation}\label{swell}
v(t)=V[1+\alpha(u(t-t_1)-u(t-t_2))]sin(\omega t)
\end{equation}
\subsection{Voltage Interruption}

When the RMS voltage drops by 90\%-100\% of its rated value for 0.5 cycles to 1 min, it is termed as a voltage interruption. They can be classified as long term ($\geq 3$ min) and short term ($\leq 3$ min) interruptions. These events are caused by faults, equipment failures, control malfunctions or operator intervention \cite{classify3,book}. Sometimes, interruptions are followed by voltage sags. It is expressed as (\ref{interrupt}) where $0.9 \leq \alpha \leq 1$,
\begin{equation}\label{interrupt}
v(t)=V[1-\alpha(u(t-t_1)-u(t-t_2))]sin(\omega t))
\end{equation}
\subsection{Voltage Harmonics}

Operation of non-linear load like variable speed drives and induction arc furnaces often distorts the sinusoidal supply voltage waveform by introducing harmonic components which are integer multiple of fundamental frequency \cite{harmonics}. These harmonics can be expressed as a sum of sinusoidal waveforms with different frequencies w.r.t the fundamental. Power factor correction capacitors also adversely increase the harmonics when distortions are already present \cite{classify3}. This results in transformer and induction motor overheating, overloaded neutrals, failure of circuit breakers and  transformers \cite{harmonic_tx}. IEEE Standard 519-2014 describes 5\% of total harmonic distortion (fraction of sum of all harmonic components compared to the RMS value at 60 Hz) as nominal. Harmonics can be mathematically described by (\ref{harmonics}) where $0.05 \leq \alpha_3, \alpha_5, \alpha_7 \leq 0.15, \sum (\alpha_i)^2=1$,\begin{equation} \label{harmonics}
v(t)=\sum_{\alpha=1,3,5,7} \alpha sin(\alpha \omega t)
\end{equation}

\subsection{Voltage Transients}

Power system transient can be described as undesirable momentary oscillations caused due to sudden change in voltage, current or load \cite{classify3}. They are broadly divided into two types: impulsive events such as lighting strokes and oscillatory events such as operation of capacitor banks or transformer start-up. Oscillatory transients can be further sub-divided into high, medium and low frequency components. Mathematically, they are described by (\ref{transients}) where $0.1 \leq \alpha \leq 0.8, w_n = 100-400Hz, \tau = 0.008-0.04$ sec,\begin{equation} \label{transients}v(t)=sin(\omega t)+\alpha e^{-(t-t_1)/\tau}sin(\omega_n t)
\end{equation}   
\subsection{Voltage Flicker}

Frequent voltage variations can be observed in appliances such as television screens or lamps which occur due to operation of arc furnaces, arc welding machines, large motors and power factor correction capacitors are termed as voltage flickers \cite{flicker}. It is often characterized by low voltage frequency between 0.5 to 25 Hz with slow magnitude changes. Voltage flickers adversely affect and reduce the span of appliances \cite{flicker2}. Traditionally, it can be modeled  as (\ref{flicker}) where $\alpha = 0.1 - 0.2$ and $ \beta = 0.5 - 25$ Hz, \begin{equation} \label{flicker}v(t)=(1+\alpha sin(\beta \omega t))sin(\omega t)
\end{equation}   

\section{Convolutional Neural Network}

In this section, we aim to classify the power quality disturbance events using 1-D convolutional neural networks. CNN has been widely used in areas of computer vision and image processing resulting in high accuracies. It is able to extract high level information from raw data eliminating dependence on domain knowledge. CNN has gained proved to be effective in classifying time series events. Authors in \cite{cnn_tim1} have used 1-D feature maps extracted from raw time series data obtained from gyroscopic sensors. Authors in \cite{cnn_tim2} have used CNN to classify time series by first converting time series to recurrence plots. This section briefly explains the various layers of CNN which is used for classification. The entire network can be expressed as a feed-forward process of cascading functions $f_k$, operating on inputs $x$ and learnable parameters $\lambda_k$ as $f(x)=f_k(....f_3(f_2(f_1(x,\lambda_1),\lambda_2,),\lambda_3),\lambda_k).$

\subsection{Input Layer}

We start with time series obtained from various power quality events of dimension $t \times 1$ with $t$ time steps. 1-D feature maps are extracted from this input over multiple CNN layers.

\subsection{Convolution Layer}
This layer performs a convolution between a portion of the time-series with learnable filters (or weights $w$). Multiple filters are used to obtain several feature maps from the inputs at each layer. These feature maps are stacked together and fed as input to the next layer. These filters resemble local receptive fields, learning from one specific subregion of the time series. Unlike traditional neural network where each weight is used once, CNN introduces the concept of parameter sharing where each filter is used at every position of the time series.

\subsection{Non-Linear Layer}

The next layer introduces non-linearity between the input and output of the convolutional layer. The result obtained after convolving 1-D feature maps with different filters is passed through a non-linear activation function. This squashes the output between a certain threshold. Non-linear activation functions such as sigmoid, rectified linear units (ReLu) or leaky-Rectified linear unit (leaky-ReLU) can be used. In this paper, we have used the LeakyReLU which keeps positive values unchanged and alleviates the problem of dying gradients in ReLu by introducing a parameter $\alpha=0.001$. It can be written as, $g_{i,j,k}=max(\alpha x,x)$.


\subsection{Fully-Connected Layer}

Feature maps generated from the previous layers are combined in this last layer which results in a $k$-dimensional vector of class probabilities. The final classification is performed using a Softmax classifier which is given as,\begin{equation}
f_k(z)=\frac{e^{z_k}}{\sum_{k}e^{z_k}}
\end{equation}Once the vector of final scores $z$, is computed by the fully-connected layer, the Softmax function normalizes it to have values within $(0,1)$. This vector can be intuitively thought as normalized class probabilities. The Softmax function can then be combined with the cross-entropy loss as,\begin{equation}
loss=-log(\frac{e^{z_k}}{\sum_{k}e^{z_k}})
\end{equation}which can be interpreted as minimizing the cross-entropy between the original and the predicted distribution. In other words, it minimizes the negative log likelihood of the correct class. 

\subsection{Parameter Updates}

In order for the network to learn, the weights can be updated using back-propagation techniques using various algorithms like gradient descent, momentum update, Nestrov momentum, AdaGrad, RMSProp, Adam and Nadam updates. Instead of training an entire batch of data which can be computationally expensive, mini-batches of data are used during each training cycle. In this paper, we have used the Nesterov-accelerated Adaptive Moment Estimation (Nadam) update mechanism, which combines RMSprop with momentum and Nesterov accelerated gradient.

\section{Results and Discussions}

Six different power quality time series: voltage sag, swell, interruption, harmonics, transients and flicker are used for classification and are given in Table \ref{tab:events}. All simulations are done using MATLAB using eqs (\ref{sag} - \ref{flicker}). The input waveform is sampled at the rate of 5KHz which is equal to the sampling rate of any industry graded PMU. In order to prevent over-fitting of the data, several methods have been used. First, the training data was split into training and validation set using 10-fold cross-validation (CV). To account for class imbalance, we have used stratified k-fold CV which takes into account the relative distribution of classes. Second, the training and the validation accuracies and losses are monitored for every epoch. The training is stopped when the validation accuracy shows no improvements above 0.0001 after 10 epochs. The average number of epochs before early-stopping was found to be roughly between 17 to 30. The training data is also randomly shuffled before each epoch. To study the robustness of the method against random measurement noise, we pollute the signals with white Gaussian noise of 80dB. The resultant classification accuracies are shown in Table \ref{tab: accuracies}. Fig. \ref{fig:confusion} presents the confusion matrix of the classification. It is seen that CNN with accuracy of 99.83\% outperforms many of the traditional approaches \cite{review}.\begin{table}[]
	\centering
	\caption{Power Quality Events}
	\label{tab:events}
	\begin{tabular}{|l|l|l|}
		\hline
		Events               & Class & Samples \\ \hline
		Voltage Sag          & 1     & 500     \\ \hline
		Voltage Swell        & 2     & 500     \\ \hline
		Voltage Interruption & 3     & 500     \\ \hline
		Voltage Harmonics    & 4     & 500     \\ \hline
		Voltage Transients   & 5     & 500     \\ \hline
		Voltage Flicker      & 6     & 500     \\ \hline
	\end{tabular}
\end{table}
\begin{table}[]
	\centering
	\caption{Accuracies}
	\label{tab: accuracies}
	\begin{tabular}{|l|c|c|l|}
		\hline
		\multirow{2}{*}{\textbf{Network}} & \multicolumn{1}{l|}{\multirow{2}{*}{\textbf{Structure}}} & \multicolumn{2}{l|}{\textbf{Accuracies}} \\ \cline{3-4} 
		& \multicolumn{1}{l|}{} & \multicolumn{1}{l|}{\textbf{Original}} & \textbf{Noisy} \\ \hline
		\textbf{CNN-1a} & \begin{tabular}[c]{@{}c@{}}3 layers, 200x1,\\ 100x1,50x1\end{tabular} & 99.83 & 99.67 \\ \hline
		\textbf{CNN-1b} & 2 layers, 200x1,100x1 & 99.52 & 98.83 \\ \hline
		\textbf{CNN-1c} & 1 layer, 200x1 & 99.81 & 99.72 \\ \hline
		\textbf{CNN-1d} & 1 layer, 400x1 &  99.79 & 99.75 \\  \hline
	\end{tabular}
\end{table}

\begin{table}[!hptb]
	\centering
	\caption{Comparison of Classification Accuracies}
	\label{tab:review}
\begin{tabular}{|l|l|l|l|l|}
	\hline
	\textbf{Classifier} & \textbf{Feature} & \textbf{Data} & \textbf{Noise} & \textbf{Accuracy} \\ \hline
	\textbf{CNN} & \textbf{High level} & \textbf{Synthetic} & \textbf{Yes} & \textbf{99.83} \\ \hline
	\textbf{Fuzzy expert} & S-Transform & Synthetic & Yes & 99.00 \\ \hline
	\textbf{Fuzzy C-means} & S-Transform & Synthetic & No & 95.41 \\ \hline
	\textbf{\begin{tabular}[c]{@{}l@{}}Multi-wavelet \\ Neural Networks\end{tabular}} & Wavelets & Synthetic & No & 98.03 \\ \hline
	\textbf{Neural Network} & Wavelets & Synthetic & Yes & 99.56 \\ \hline
	\textbf{SVM} & Wavelets & Practical & Yes & 95.81 \\ \hline
	\textbf{k-NN} & Hilbert, Clarke & Synthetic & No & 80.06 \\ \hline
	\textbf{Hidden Markov} & Fourier,Wavelet & Synthetic & No & 95.71 \\ \hline
	\textbf{\begin{tabular}[c]{@{}l@{}}Decision Tree \\ Fuzzy\end{tabular}} & S-Trasform & Practical & Yes & 97.56 \\ \hline
	\textbf{\begin{tabular}[c]{@{}l@{}}Radial Basis \\ Function\end{tabular}} & Wavelet & Synthetic & Yes & 96.6 \\ \hline
\end{tabular}
\end{table}

\begin{figure}[!hptb]
\centering
\includegraphics[scale=0.31]{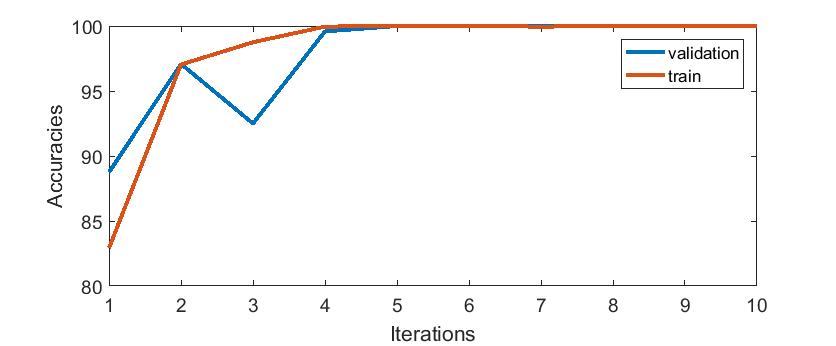}
\caption{Training and Validation Accuracies for nosiy singal}
\label{fig:accuracies}
\end{figure}
\begin{figure}[!hptb]
	\centering
	\includegraphics[scale=0.48]{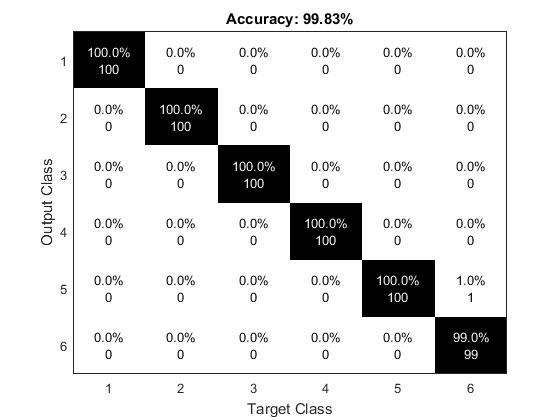}
	\caption{Confusion Matrix: Original Signals}
	\label{fig:confusion}
\end{figure}\begin{figure}[!hptb]
\centering
\includegraphics[scale=0.48]{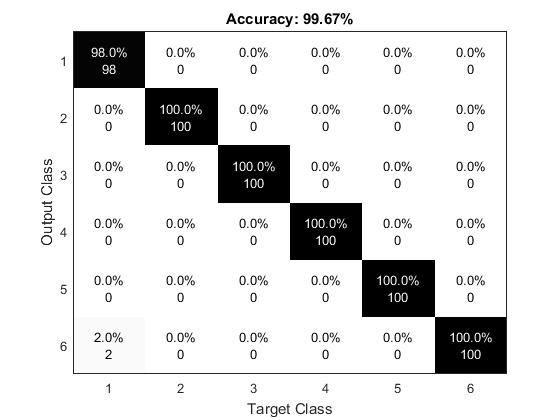}
\caption{Confusion Matrix: Original Signals + 80 dB WGN}
\label{fig:confusion}
\end{figure}


\section{Future Works}
To improve this paper, here are the following things that still need to be done.

\begin{enumerate}
	\item Generate more power quality signals such as combination of sag and swell with harmonics, notch, spikes,.
	\item Use RTDS to generate power system fault waveforms and subsequently use it to test accuracy of model obtained from simulation data.
	\item Classify other events such as induction motor start-up, self-extinguishing faults, transformer and line energising. 

\end{enumerate}

\ifCLASSOPTIONcaptionsoff
  \newpage
\fi


\begin{thebibliography}{1}
\bibitem{intro1}Saleh, S. A. "Phaselet Transform-Based Approach for Detecting Voltage Flickers Due to Distributed Generation Units." IEEE Transactions on Industry Applications (2017).

\bibitem{intro2}Kumar, Raj, Bhim Singh, and D. T. Shahani. "Symmetrical components-based modified technique for power-quality disturbances detection and classification." IEEE Transactions on Industry Applications 52, no. 4 (2016): 3443-3450.

\bibitem{lit1}Singh, Utkarsh, and Shyam Narain Singh. "Application of fractional Fourier transform for classification of power quality disturbances." IET Science, Measurement and Technology 11, no. 1 (2017): 67-76.

\bibitem{lit2}Lee, Chun-Yao, and Yi-Xing Shen. "Optimal feature selection for power-quality disturbances classification." IEEE Transactions on power delivery 26, no. 4 (2011): 2342-2351.

\bibitem{powerQ_CNN}Binsha P., Sachin Kumar S., Athira S. and K. P. Soman, `Power Quality Signal Classification using Convolutional Neural Network.'International Journal of Computer Technology and Applications, 2016, pp. 8033-8042

\bibitem{classifiy1} Erişti, H., and Y. Demir. "Automatic classification of power quality events and disturbances using wavelet transform and support vector machines." IET generation, transmission and distribution 6, no. 10 (2012): 968-976.

\bibitem{classify2}Ekici, Sami. "Classification of power system disturbances using support vector machines." Expert Systems with Applications 36, no. 6 (2009): 9859-9868.

\bibitem{classify3}Chintakindi, Sruthi Reddy, and DVSS Siva Sarma. "Classification of Power Quality events using improved S-transform." In India Conference (INDICON), 2015 Annual IEEE, pp. 1-5. IEEE, 2015.

\bibitem{book}Dugan, Roger C., Mark F. McGranaghan, and H. Wayne Beaty. "Electrical power systems quality." New York, NY: McGraw-Hill,| c1996 (1996).

\bibitem{harmonics}Phipps, James K., John P. Nelson, and Pankaj K. Sen. "Power quality and harmonic distortion on distribution systems." IEEE transactions on industry applications 30, no. 2 (1994): 476-484

\bibitem{harmonic_tx}Henderson, Robert D., and Patrick J. Rose. "Harmonics: The effects on power quality and transformers." IEEE transactions on industry applications 30, no. 3 (1994): 528-532.

\bibitem{flicker}Sen, Priyanka., Panda, K.P., R, Sreyasee, "Enhancement of Power Quality and Voltage Flicker Mitigation Using New PWM Based DSTATCOM."2017 International Conference on Intelligent Computing and Control Systems (ICICCS), 2017. 

\bibitem{flicker2}Saleh, S. A. "Phaselet Transform-Based Approach for Detecting Voltage Flickers Due to Distributed Generation Units." IEEE Transactions on Industry Applications (2017).

\bibitem{cnn_tim1}Zhao, Bendong, Huanzhang Lu, Shangfeng Chen, Junliang Liu, and Dongya Wu. "Convolutional neural networks for time series classification." Journal of Systems Engineering and Electronics 28, no. 1 (2017): 162-169. 

\bibitem{cnn_tim2}Hatami, Nima, Yann Gavet, and Johan Debayle. "Classification of Time-Series Images Using Deep Convolutional Neural Networks." arXiv preprint arXiv:1710.00886 (2017).

\bibitem{review}Saini, Manish Kumar, and Rajiv Kapoor. "Classification of power quality events–a review." International Journal of Electrical Power and Energy Systems 43, no. 1 (2012): 11-19.s



\end{thebibliography}
\end{document}